# Transmission electron goniometry and its relation to electron tomography for materials science applications

Peter Moeck
Nanocrystallography Group, Department of Physics, Portland State University, P.O. Box 751, Portland, OR 97207-0751, U.S.A, Oregon Nanoscience and Microtechnologies Institute, http://www.onami.us

Philip Fraundorf
Department of Physics and Astronomy and Center for Molecular Electronics, University of Missouri at St. Louis, MO 53121, U.S.A.

## ABSTRACT

Aspects of transmission electron goniometry are discussed. Combined with high resolution phase contrast transmission electron microscopy (HRTEM) and atomic resolution scanning TEM (STEM) in the atomic number contrast (Z-STEM) or the phase contrast bright field mode, transmission electron goniometry offers the opportunity to develop dedicated methods for the crystallographic characterization of nanocrystals in three dimensions. The relationship between transmission electron goniometry and electron tomography for materials science applications is briefly discussed.

Internet based java applets that facilitate the application of transmission electron goniometry for cubic crystals with calibrated tilt-rotation and double-tilt specimen holders/goniometers are mentioned. The so called "cubic-minimalistic" tilt procedure for the determination of the lattice parameters of sub-stoichiometric $WC_{1-x}$ nanocrystals with halite structure is demonstrated.

The open-access, Internet based, Crystallography Open Database (COD) and its mainly inorganic subset are briefly discussed because the ability to determine the lattice parameters of nanocrystals opens up the possibility of identifying unknown crystal phases by comparing these lattice constants to the entries of crystallographic databases. "Lattice-fringe fingerprinting in two dimensions" for the identification of unknown phases is mentioned in passing as a simple alternative (that does not involve the usage of specimen goniometers). The enhanced viability of transmission electron goniometry in microscopes with aberration corrected electron optics is illustrated on the $WC_{1-x}$ model system.

## INTRODUCTION

Phase diagrams and crystal morphologies are both crucial to all kinds of properties of nanoparticles and frequently depend on the size of the crystals in the nanometer range. Added to this size dependency of the lowest thermodynamical potential of a structure, there is in the nanocrystal regime a strong tendency to metastability and spatially inhomogeneous non-stoichiometry. Many nano-crystals possess structural defects in three (3D), two (2D) and one (1D) dimension(s) as well as point defects well in excess of their thermodynamical equilibrium levels. These structural defects affect the properties of the nanocrystals and are strongly dependent on the particulars of the nanocrystal synthesis and processing procedures. In short, a whole new "crystallographic world" is waiting to be discovered (and later on employed) in the nanocrystal realm.

The aim of this paper is to discuss aspects of transmission electron goniometry. The relationship between transmission electron goniometry and electron tomography for materials science applications is also discussed briefly. We mention internet based java applets that can be employed for the transmission electron goniometry of cubic crystals with calibrated tilt-rotation and double-tilt TEM (or STEM*) specimen holders/goniometers. We discuss the so called "cubic-minimalistic" tilt procedure for the determination of the lattice parameters of sub-stoichiometric $WC_{1-x}$ nanocrystals with halite structure.

Since the ability to determine the lattice parameters of nanocrystals opens up the possibility of identifying unknown phases by comparing these lattice constants to the entries of crystallographic databases, we briefly discuss the Internet based, open-access Crystallography Open Database (COD) and its mainly inorganic subset.

The enhanced viability of image-based nanocrystallography by means of transmission electron goniometry in TEMs (or STEMs) with aberration-corrected electron optics [1,2] is illustrated in the final section of this paper. Harald Rose's prediction from more than ten years ago: *"In materials science tomographic methods will become important at a resolution limit which allows one to use different crystal orientations."* [1] might well be considered as beginning to be fulfilled with the emergence of image-based nanocrystallography in three dimensions by means of transmission electron goniometry.

Although discussed here mainly in connection with high resolution phase contrast imaging, transmission electron goniometry can, as such, be employed under either quasi-kinematical (phase-contrast imaging, parallel bean electron diffraction to derive structure factor amplitudes, …), incoherent (Z-contrast imaging), or dynamic (Kikuchi diffraction, convergent beam electron diffraction, …) conditions.

## TRANSMISSION ELECTRON GONIOMETRY

Two complementary versions of transmission electron goniometry have been proposed by the authors of this paper [3,4]. P. Fraundorf's method [3] can be considered as constituting goniometry of reciprocal lattice vectors on the basis of Fourier transform power spectra of series of high-resolution electron phase-contrast images. P. Moeck's method [4], on the other hand, constitutes goniometry of direct lattice vectors and has so far been mainly used in connection with series of Kikuchi-pole patterns for direct crystallographic analyses, e.g. orientation determinations, orientation relationship determinations [5], texture estimations [6], and grain- and phase boundary parameter determinations [7]. A combination [8] of the two transmission electron goniometry versions is possible and desirable for increased utility and accuracy of image-based nanocrystallography.

Goniometry is described by various dictionaries as "defining and interrelating of the functions of angles to solve practical problems". It means in the context of this paper that a TEM (or STEM) specimen holder that is calibrated and possesses a tilt/readout accuracy of at least ± 1° (commonly defined as a goniometer [9]) is employed to derive the coordinates of reciprocal [3,8,10,11] and/or direct lattice vectors [4-7] in a Cartesian coordinate system that is fixed to the electron microscope. Regardless of which type of calibrated specimen holder is employed, there is always a set of simple trigonometric relations that lead from the curvilinear coordinates (goniometer setting readings) of this goniometer to the Cartesian coordinates of the (direct and reciprocal) lattice vectors that are visible in images.

Indexing these lattice vectors also in their crystallographic coordinate system then provides the basis to calculate a 3 by 3 transformation matrix between the crystallographic coordinate system of the nanocrystal under investigation and the Cartesian coordinate system of the electron microscope. This 3 by 3 matrix is frequently called the crystal matrix (or structure matrix [12]) of either the direct or reciprocal lattice and lends itself perfectly to all kinds of crystallographic analyses. The metric tensor is, e.g., obtained by multiplying the crystal matrix with its transposed matrix from the left. From the components of the metric tensor, the lattice parameters can be easily calculated.

The transformation / crystal / structure matrices represent the ideal crystal structure at the lattice point level. While structural 3D defects can be represented by their own crystal matrices, 2D and 1D defects can be described in the framework of the crystal matrices after they have been amended with space group information [13].

Note that the above description was completely general. Any symmetry the crystal under investigation may possess is not taken advantage of in the calculations. The described procedure is, therefore, applicable to any type of crystal regardless of the symmetry of its atomic arrangement.

It is straightforward to see that the crystallographic analysis techniques that have been developed on the basis of optical goniometry (i.e. classical crystallometry, also called morphological crystallography) [14], the stereographic projection [15], and transmission light goniometry [16] can be used in modified form for transmission electron goniometry in TEMs and STEMs. Recently, a java-based applet has been described that facilitates the usage of the stereographic projection for transmission electron goniometry [17]. Java-based applets that simulate a tilt-rotation TEM goniometer [18] and a double-tilt TEM goniometer [19] are accessible freely over the internet.

Within their angular range, conventional TEM (or STEM) specimen holders with two axes can be used to orient a crystal in any orientation with respect to the electron beam. Transmission electron goniometry can, thus, be practiced with calibrated two-axis specimen holders of the current designs. The angular range of a typical ± 20° tilt 360° rotation holder is nine times larger that that of a typical ± 20° double-tilt holder. In addition, it has the advantage that a specific zone axis or net plane normal about which one wants the tilt the crystal can be adjusted parallel to the eucentric axis of a side-entry holder. The double-tilt goniometer, on the other hand, has been made *compu*-centric by commercially available software (i.e. compu-centricity for FEI tecnai microscopes with compustage). A double-tilt TEM goniometer with an extra degree of freedom to rotate the specimen by 360° is commercially available from Gatan Inc. and allows electron microscopists to harness the well established procedures of classical crystallometry [14] for all kinds of image-based nanocrystallographic analyses in three dimensions.

Surely, with space for specimen goniometers in the centimeter range in all three dimensions that aberration-corrected optics will eventually deliver, future goniometers could be made nearly eucentric by novel mechanical designs and fully *compu*-centric by computer control. It should be noted briefly that there is precedence from the late 80s of the combined design of both a high resolution objective lens and an eucentric ± 45° and ± 23° double-tilt goniometer [20]. So eucentric specimen goniometers are indeed possible provided that there is sufficient space in the microscope column close to the objective lens.

In addition, when aberration corrected electron optics [1,2] are used, high precision crystal tilts can be replaced by even higher precision electron beam tilts. Since such replacements are complementary, transmission electron goniometry procedures are unaffected conceptually, lessening the demands on specimen goniometers for precision crystal tilts.

The most versatile TEM specimen holder would be (*compu-*)eucentric and allow for two mutually perpendicular tilts and a perpendicular rotation. If this holder is calibrated in all three axes and sufficiently accurate tilts/readouts are provided [9], it provides basically the same functionality as a goniometer head, Fig. 1, which is fixed onto the turn table of a single-circle optical goniometer.

While two tilt/rotation axes are (within their range) sufficient to orient any crystal direction parallel to any orientation in space, e.g. the optical axis of a microscope or the third translation axis, the third tilt/rotation axis of such a goniometer allows the oriented crystal to be rotated around this orientation in space.

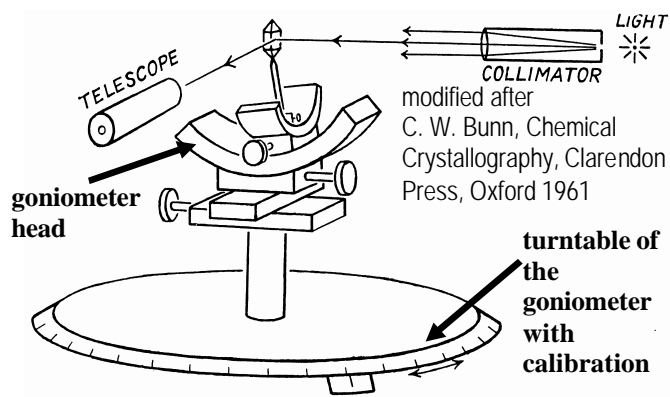

modified after
C. W. Bunn, Chemical Crystallography, Clarendon Press, Oxford 1961

**Figure 1:** Goniometer head fixed onto the turn table of a single-circle optical goniometer. The direction of the light source is fixed, just as the direction of the electron beam in a TEM (or STEM). Since the crystal is not transmitted, the "detector" is not placed opposite to the light source, as indicated by the light ray arrows. One crystal face is reflecting at a time and the normals of the reflecting crystal faces are in a fixed direction in space. There are three degrees of freedom to tilt the crystal and three degrees of freedom to translate the crystal.

## RELATION OF TRANSMISSION ELECTRON GONIOMETRY TO ELECTRON TOMOGRAPHY FOR MATERIALS SCIENCE APPLICATIONS

First it needs to be clarified what is generally meant by electron tomography. The introduction to a widely quoted multi-author text defines electron tomography as *"any technique that employs the transmission electron microscope to collect projections of an object and uses these projections to reconstruct the object in its entirety. Excluded from this definition, more or less for historical reasons, are techniques that make use of inherent order or symmetry properties of the object."* [21].

Such excluded techniques have been pioneered in the late 60s by Aaron Klug (Nobel Prize 1983 *"for his development of crystallographic electron microscopy and his structural elucidation of biologically important nucleic acid-protein complexes"*) and coworkers [22], whereby the objects of their studies possessed either non-translational periodic, e.g. helical or icosahedral, symmetries, or where crystalline in two dimensions. As crystallographic concepts are applied, an appropriate term for these techniques is *"crystallographic electron tomography"*, which

Aaron Klug used himself.

Many kinds of tomography (i.e. reconstruction of a 3D object from a series of its 2D projections) are based on the inverse Radon transform and Radon's projection theorem [23-25]. For electron tomography this requires that the individual 2D electron tomograms (images) should be dominated by "projection contrast" or "scattering contrast" [26] which means that the recorded electron intensity must at least be a monotonic function of the thickness or density of the 3D object. Ideally Bear's law applies and the numbers of electrons per pixel areas of a recording charged coupled device camera are proportional to $e^{-\mu t}$ (with $\mu$ as linear absorption coefficients and t as thicknesses of the object in the transmitted direction). Then the logarithms of these numbers represent the projection through the object. Stained and unstained amorphous biological objects frequently fulfill this requirement [26].

Most specimens of interest in materials science are, however, crystalline and their images in conventional (parallel illumination) TEMs are dominated by interference effects such as phase contrast, diffraction contrast, dynamical amplitude contrast phenomena (i.e. bend contours and thickness fringes), and Fresnel fringes at the specimen edges [25,27].

Energy-filtered imaging in TEM allows for a suppression of interference effects and has been employed for electron tomography [28], but these images frequently possess a poor signal-to-noise ratio. Z-STEM images on the other hand, may possess a better signal-to-noise ratio, and arise from essentially incoherent Rutherford scattering so that interference effects which violate the projection theorem are negligible. Such (high angle annular dark field) images are, therefore, preferentially used for electron tomography [25,29-31].

Typically some 100 2D tomograms are needed for the reconstruction of a 3D object with a volume of approximately 100 nm$^3$ [25]. The spatial resolution of STEM images can in contemporary TEM/STEMs be as high as approximately 0.2 nm, but the volume resolution of a reconstructed 3D object that is visualized by (quasi-continuous) Z-STEM tomography is at best on the order of 1 nm$^3$ [29,30], i.e. much coarser than the atomistic structure of crystals.

These 100 or so images are taken quasi-continuously (i.e. every few degrees) over as large a tilt range as it is practicable. Since images are for a variety of reasons hardly ever recorded over the full ± 90° tilt range, there is typically object information missing in the 3D reconstruction [24,25,30]. This missing object information results in blurring of the 3D reconstruction, elongation of specimen features in the direction parallel to the optical axis of the microscope, and other artifacts such as, e.g. occasionally an apparently missing arm [31] of nanometer sized CdSe tetrapods with zinc blende core and four wurtzite arms [32]. When the specimen contains faces, the interface angles are accordingly distorted.

There is a strong electron channeling effect in Z-STEM images whenever a crystal is transmitted along a low indexed zone axis. Z-STEM images of such low indexed zone axis are naturally contained in most 2D electron tomogram sets, but must be discarded before the 3D reconstruction of the object is undertaken since their intensity is much higher than that of the other tomograms in the set.

This fact provides an insight into the relation between standard (quasi-continuous) electron tomography and 3D image-based nanocrystallography on the basis of transmission electron goniometry. It is exactly those low-indexed zone axis Z-contrast (or electron phase contrast) images that standard electron tomography has to discard which form the backbone of image-based nanocrystallography in three dimensions.

Typical 3D image-based nanocrystallography methods are discrete atomic resolution tomography [33] and atomic structure determination from discrete sets of HRTEM images [34-39] (electron crystallography) which were recorded at selected crystallographic orientations. Since crystals [33-35], quasi-crystal approximants [36] or quasi-crystals [37] have been analyzed, only a few, but yet very characteristic 2D projections have been employed in all of these studies. Pre-selecting such 2D projections on theoretical grounds and actually supporting the tilting of the nanocrystals into such 3D orientations are straightforward applications of transmission electron goniometry.

When discrete atomic resolution tomography is performed on the basis of transmission electron goniometry, there is no missing object information and the interface angles of nanocrystals are not distorted. When the crystal matrix is known, the indices of the faces can be straightforwardly calculated from the cross products of direct lattice vectors that lie within a face as seen in the discrete set of 2D projections. This is akin to classical trace analysis [15] from two 2D projections.

# CUBIC-MINIMALISTIC TILT STRATEGIES FOR DETERMINING LATTICE PARAMETERS

The "cubic-minimalistic" tilt strategies [11] for determining the lattice parameters of individual cubic nanocrystals from only two high-resolution phase-contrast images present an opportunity to practice image-based nanocrystallography with contemporary high resolution TEMs with a modest tilt range at a minimal tilting effort. We define here a TEM which a point-to-point resolution of 0.2 nm or better as a HRTEM.

The tilt procedure is in one of its basic forms, which relies in this particular case on the prior knowledge that the nanocrystals belong to the space group $Fm\bar{3}m$, schematically shown in Fig. 2 Other cubic minimalistic tilt strategies (for nanocrystals that belong to other cubic space groups) and the determination of the lattice constants of approximately 10 nm wide sub-stoichiometric $WC_{1-x}$ nanocrystals with the halite structure are described in detail in ref. [11].

We illustrate below briefly how this determination can be achieved by goniometry of direct lattice vectors combined with the measurement of the visible lattice fringe spacings, the determination of the crystal matrix, and the metric tensor. An ensemble of dispersed $WC_{1-x}$ nanocrystals were imaged at the two goniometer settings of a double-tilt TEM holder: $a_1$ (or eucentric tilt): $15 \pm 0.1°$; $\beta_1$ (or non-eucentric tilt): $9.7 \pm 0.1°$; and $a_2$: $-15 \pm 0.1°$, $\beta_2$: $-9.7 \pm 0.1°$. These two goniometer settings correspond to a single counterclockwise tilt of approximately $35.3°$ about an effective tilt axis that is perpendicular to the electron beam.

It was kept track of the movement of the individual nanocrystals. Those nanocrystals that showed crossed {200} lattice fringes in one image and a corresponding single set of (11-1) lattice fringes in matching orientation in the second image were selected for careful analysis. For this combination of lattice fringes, the effective tilt axis coincides with the [110] axis of the nanocrystals.

The following is mainly intended to demonstrate the intrinsic connection between goniometry of direct and reciprocal lattice vectors. Employing goniometry of direct lattice vectors to the same data set one calculates from these goniometer settings and the corresponding zone axes, the following normalized crystal matrix of the direct lattice:

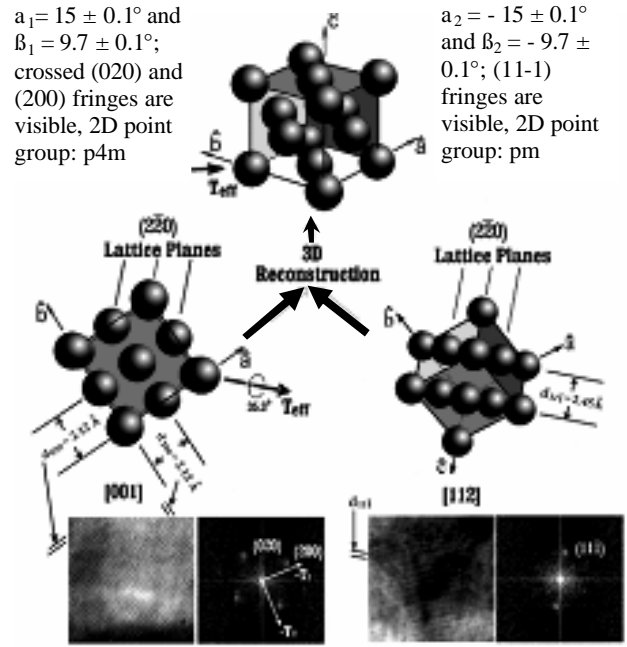

$a_1 = 15 \pm 0.1°$ and $\beta_1 = 9.7 \pm 0.1°$; crossed (020) and (200) fringes are visible, 2D point group: p4m

$a_2 = -15 \pm 0.1°$ and $\beta_2 = -9.7 \pm 0.1°$; (11-1) fringes are visible, 2D point group: pm

**Fig. 2:** Sketch of the "cubic-minimalistic" tilt procedure for a sub-stoichiometric $WC_{1-x}$ nanocrystal with the halite structure. Only the W atoms are schematically shown as they are about twice as large as the C atoms. The C atoms are supposed to fill the octahedral intersites in the face centered cubic W-sublattice more or less regularly. The Scherzer resolution of the employed TEM was 0.19 nm. One would need a directly interpretable resolution of at least 0.15 nm to resolve the {220} fringes so that the [112] zone axis would be revealed by crossed lattice fringes as well. (Modified on the basis of a figure from the PhD thesis of Wentao Qin.)

$$\begin{pmatrix} -0.2337 \pm 0.004 & 0.9585 \pm 0.0029 & -0.1629 \pm 0.0016 \\ -0.9484 \pm 0.004 & -0.1878 \pm 0.0026 & 0.2554 \pm 0.0016 \\ 0.2142 \pm 0.0002 & 0.2142 \pm 0.00015 & 0.9530 \pm 0.0007 \end{pmatrix}$$

From this matrix and the measured lattice fringe spacings, one obtains the metric tensor of the direct lattice in $nm^2$:

$$\begin{pmatrix} 0.180622 \pm 2.5 \cdot 10^{-5} & (0.41 \pm 2.2) \cdot 10^{-5} & (0.26 \pm 2.1) \cdot 10^{-5} \\ (0.41 \pm 2.2) \cdot 10^{-5} & 0.180620 \pm 2.1 \cdot 10^{-5} & (0.31 \pm 2.0) \cdot 10^{-5} \\ (0.26 \pm 2.1) \cdot 10^{-5} & (0.31 \pm 2.0) \cdot 10^{-5} & 0.180623 \pm 1.9 \cdot 10^{-5} \end{pmatrix}$$

From this tensor, within experimental errors, a cubic lattice with a constant of $0.425 \pm 0.005$ nm is obtained. The transmission electron goniometry procedure leads in this case to smaller errors (~ 0.2 %) than the determination of the lattice constant from the two images (~ 1 %). When crystallographic image processing [34-39] is employed (to more or less perfectly periodic regions of the specimen only), the crystallographic accuracy and precision of the effective tilt between the images can be further increased, resulting in even smaller experimental errors due to the transmission electron goniometry procedure. Imposing the projected 2D symmetry of the electrostatic potential in HRTEM zone axis images by means of crystallographic image processing [37,38], thus, lessens the demands on the specimen goniometer.

Note that while lattice constant determinations from electron phase contrast lattice fringes for 5 to 15 nm sized nanocrystals that contain 2D and 3D defects are typically accurate to within approximately 0.5 %, for structurally perfect nanocrystals a relative precision of 0.01 % and an absolute accuracy of 0.005 nm is obtainable [40-42]. Larger tilt angles would be required for lower symmetric structures, but improvements of the directly interpretable image (point-to-point microscope) resolution that aberration corrected electron optics provide reduce these tilt angles super-linearly as there is a strongly super-linear increase of the visibility of lattice fringes and zone axes with an increasing directly interpretable image resolution [43], Table 1.

The modest tilt range of current TEM specimen goniometers is, therefore, tolerable when lens aberration correction improves the directly interpretable image resolution down to below 0.1 nm. Large angle TEM specimen goniometers that are possible with such instruments will improve the accuracy with which the lattice vector components parallel to the electron beam can be measured [3] and, therefore, increase the accuracy of all kinds of crystallographic analyses on the basis of transmission electron goniometry.

It is noted in passing that well developed cryogenic 3D electron microscopy reconstruction techniques such as angular reconstitution and random conical tilt reconstruction** can also be adopted for the development of novel image-based nanocrystallography techniques. While it is in structural biology rather unimportant which tilt angle is used in a random conical tilt reconstruction, in image-based nanocrystallography this tilt angle needs to be part of a tilt strategy.

## OPEN-ACCESS CRYSTALLOGRAPHIC DATABASE SUPPORT

Being able to determine the lattice parameters of nanocrystals opens up the possibility of identifying unknown phases by comparing these parameters to those of the entries of crystallographic databases [44]. Good choices for such databases are the Crystallography Open Database (COD) and its mainly inorganic subset [45-47], Fig. 3, since they are "quite comprehensive", rapidly growing, and freely accessible over the Internet. A search for entries in the latter database for the experimentally determined cubic lattice vector magnitude 0.425 ± 0.005 nm returns currently 105 entries that match the search criteria. Information on the space group and the atomic coordinates are included for all entries of the COD's mainly inorganic subset and it is also possible to visualize and compare candidate structures interactively in three dimensions.

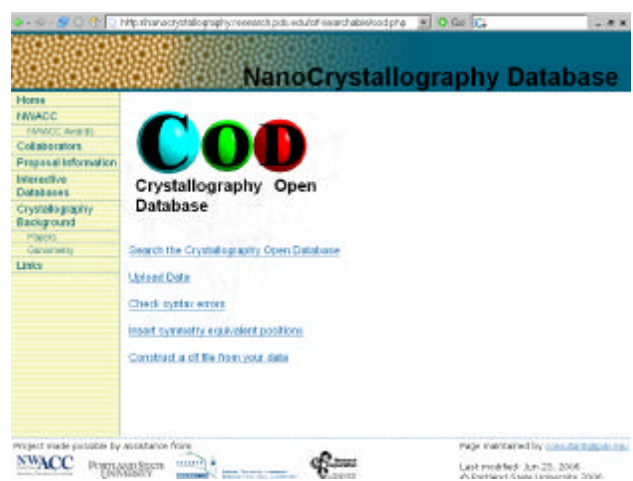

**Figure 3:** Home page of the mainly inorganic subset of the COD [47]. For candidate structures, searches can be performed with Boolean constraints for elements absent and/or present and the strict number of elements present.

It is noted in passing that there is a simple alternative to identifying unknown crystal phases from high-resolution TEM images which does not involve the usage of the specimen goniometer. We call that other method "lattice-fringe fingerprinting in 2D" [10,48,49] and note that it is also supported by the mainly inorganic subset of the COD [47].

With approximately one fifth of all know crystal structures already in the public domain and freely accessible over the Internet [44], the open-access movement is well supported by the crystallographic community. With that kind of Internet database support for the identification of unknown crystal phases available, we are now one step closer to the realization of De Ruijter's and co-workers' closing statement from their 1995 paper [42]:

*"… an obvious step in identification would be to establish an automated link with these databases and associated search/matching programs thus enabling immediate printouts of a list of possible materials and phase names."*

# PROSPECTIVE VIABILITY OF IMAGE-BASED NANOCRYSTALLOGRAPHY

Image-based nanocrystallography by means of transmission electron goniometry will be the more viable the more direct lattice vectors are accessible by tilting of the nanocrystal and the more reciprocal lattice vectors are visible in high-resolution phase-contrast TEM or atomic resolution STEM images [12], [43], and Table 1.

As mentioned above, being able to see and analyze more (direct and reciprocal) lattice vectors in the same unit of orientation space lessens the necessary angular range of the specimen goniometer, Table 2.

| Directly interpretable image resolution [nm] | Visible lattice plane types within one stereographic triangle [001]-[011]-[111] | Visible zone axes (lattice fringe crosses) within one stereographic triangle [001]-[011]-[111] |
|---|---|---|
| 0.2 | 2, i.e. {111}, {200} | 2, i.e. [001], [011] |
| 0.15 | 3, i.e. {111}, {200}, {220} | $2^2$, i.e. [001], [011], [111], [112] |
| 0.1 | 4, i.e. {111}, {200}, {220}, {311} | $2^3$, i.e. [001], [011], [111], [112], [013], [114], [125], [233] |
| = 0.05 | = 18, i.e. {111}, {200}, {220}, {311}, {331}, {420}, {422}, {511}, {531}, {442}, {620}, {622}, {551}, {711}, {640}, {642}, {731}, {820} | > $2^5$, e.g. [001], [011], [111], [012], [112], [013], [122], [113], [114], [123], [015], [133], [125], [233], [116], [134], [035], ... |

**Table 1**: Visible lattice planes and zone axes as a function of directly interpretable image resolution for sub-stoichiometric $WC_{1-x}$ nanocrystals. When we take, in a purely phenomenological manner, the number of zone axes in one stereographic triangle, column 3, as measure of the viability of image-based nanocrystallography by means of transmission electron goniometry, there is a strongly super-linear increase in viability with improvements in directly interpretable image resolution.

Many two zone-axis tilt protocols [11] and more elaborate tilt procedures than the one presented above and correspondingly more accurate and comprehensive image-based nanocrystallographic analyses particularly for unknown structures of arbitrary symmetry become possible with improved directly interpretable image resolution. With point-to-point resolutions below 0.1 nm, much more than the minimal data set for transmission electron goniometry will be accessible for basically all crystals. The crystallographic analyses can then be performed by least-squares fits to all of the experimental data. To perform a particular image-based nanocrystallographic analysis, specific tilt procedures can be designed that provide, for example, maximal accuracy for certain parameters at a reasonable effort.

| Directly interpretable image resolution [nm] | Average angle between zone axis pairs (corresponding to number of possible two zone-axis tilt protocols) | Minimum double-tilt range requirement to achieve average angle between zone axis pairs |
|---|---|---|
| 0.2 | 50° (out of the 3 zone axis pairs/tilt protocols in two stereographic triangles, [001]-[011]-[111]-[101]) | ± 18.4° |
| 0.15 | 36.6° (out of all 6 zone axis pairs/tilt protocols in one stereographic triangle) | ± 13° |
| 0.1 | 26.2° (out of all 28 zone axis pairs/tilt protocols in one stereographic triangle) | ± 9.3° |
| = 0.05 | = 9.3° (out of only those 21 zone axis pairs/tilt protocols with [u + v + w] = 8 in one stereographic triangle that are also along {111}, {200}, and {220} bands) | = ± 3.3° (when aiming only for those zone axes/tilt protocols with [u + v + w] = 8 that are also along {111}, {200}, and {220} bands) |

**Table 2**: Average angle between zone axis pairs and minimal required double-tilt range as a function of the directly interpretable image resolution for sub-stoichiometric $WC_{1-x}$ nanocrystals. When we take, again in a purely phenomenological manner, the minimum double-tilt range requirement to achieve the average angle between zone axis pairs, column 3, as measure of the viability of image-based nanocrystallography by means of transmission electron goniometry, there is again a strongly super-linear increase in viability with improvements in directly interpretable image resolution.

With support from both comprehensive databases such as the COD and aberration corrected electron microscopes [1,2], image-based nanocrystallography in three dimensions may become one of the realizations of Boldyrew's and Doliwo-Dobrowolsky's 70 years old prophecy that: *"In the further development of crystallography one will either adopt one of the goniometric methods of determining crystals or develop eventually a new one which, as far as this is possible, combines the advantages of all of the prior methods and avoids their disadvantages."* [50].

## CONCLUSIONS

Transmission electron goniometry and its relation to standard (quasi-continuous) electron tomography for materials science applications have been discussed. The metric tensor and lattice parameters have been determined for a cubic nanocrystal model system. The method is applicable to all crystals regardless of their symmetry and conceptually similar to classical crystallometry. Although feasible for cubic nanocrystals with lattice constants larger than 0.4 nm in the current generation of uncorrected high resolution TEMs with modest tilt range, image-based nanocrystallography will become much more viable in transmission electron microscopes with aberration-corrected electron optics. Comprehensive crystallographic databases such as the Crystallography Open Database and its mainly inorganic subset, which are freely accessible over the Internet, may support transmission electron goniometry and image-based nanocrystallography in three dimensions as well.

## ACKNOWLEDGMENTS

This research was supported by an award from Research Corporation. Additional support was provided by faculty enhancement and development awards from Portland State University.

---

\* STEM images typically suffer from scan aberrations such as distortions in the resolved spatial frequencies, resulting in distortions in the angles between lattice fringes. These distortions can, in principle, be corrected for each individual STEM operated under its typical imaging conditions, see, e.g., A.M. Sanchez et al., An approach to the systematic distortion correction in aberration-corrected HAADF images, *J. Microsc.* **221** (2006) 1-7.

\*\* For angular reconstitution see: e.g., M. Van Heel, Angular reconstitution: a posteriori assignment of projection directions for 3D reconstructions, *Ultramicroscopy* **21** (1987) 111-124, for random conical tilt reconstruction, see, e.g., A.J. Koster et al., Perspectives of Molecular and Cellular Electron Tomography, *J. Struct. Biolog.* **120** (1997) 276-308.